\documentclass[12pt]{article}
  \usepackage{epsfig,amsfonts}
  \usepackage{axodraw}
  \usepackage{graphicx}
  \usepackage{subfigure}
\evensidemargin - 1.truecm
\def\be{\begin{equation}}
\def\ee{\end{equation}}
\def\bc{\begin{center}}
\def\ec{\end{center}}
\def\bfig{\begin{figure}}
\def\efig{\end{figure}}
\def\bea{\begin{eqnarray}}
\def\eea{\end{eqnarray}}
\def\dd{\displaystyle}
\def\nn{\nonumber}
\begin{document}

\def\be{\begin{equation}}
\def\ee{\end{equation}}
\def\bc{\begin{center}}
\def\ec{\end{center}}
\def\bfig{\begin{figure}}
\def\efig{\end{figure}}
\def\bea{\begin{eqnarray}}
\def\eea{\end{eqnarray}}
\def\dd{\displaystyle}
\def\nn{\nonumber}

\begin{titlepage}

\vspace*{50mm}

\begin{center}
\boldmath
{\Large \bf Analytical Calculation \\[2mm] 
of Two-Loop Feynman Diagrams }\unboldmath
\vskip 1.2cm
{\large  R.~Bonciani\footnote{Email: 
{\tt Roberto.Bonciani@physik.uni-freiburg.de}}
\footnote{Presented at the final meeting of the European Network
``Physics at Colliders'', Montpellier, September 26-27 2004}}
\vskip .6cm
{\it Fakult\"at f\"ur Mathematik und Physik, Albert-Ludwigs-Universit\"at
Freiburg, \\ D-79104 Freiburg, Germany} 
\vskip .3cm
\end{center}
\vskip 2.4cm

\begin{abstract} 
We review the Laporta algorithm for the reduction of scalar integrals to
the master integrals and the differential equations technique for their
evaluation. We discuss the use of the basis of harmonic polylogarithms for
the analytical expression of the results and some generalization of this
basis to wider sets of transcendental functions.
\end{abstract}
\vfill
\end{titlepage}

\section{Introduction}

In the last few years a large amount of work has been devoted to the
improvement of the techniques for the calculation of Feynman diagrams.
The reason is that future high energy physics experiments will reach
a measurement precision that will require, from the theoretical
counterpart, the control on the NNLO quantum corrections for several 
physical observables.

Basically two approaches have been developed for the calculation 
of Feynman diagrams: the first one is based on the numerical and the other 
on the analytical evaluation of the integrals involved. The goal of both 
approaches is the complete and ``automatic'' evaluation of Feynman diagrams 
in multi-scale processes, but, nowadays, this goal is far from being achieved.
Different problems arise. While for the numerical approach the presence of
different scales is not a problem, the treatment of infrared 
singularities, thresholds and pseudo-thresholds is of complicated solution 
and sometimes it has to be performed in a semi-analytic way. On the other 
hand, the analytic approach gives a complete control on the ``difficult''
regions of the spectrum, but it is, at the moment, constrained to processes 
in which the scales in the game are at most three. 

Nevertheless, in both cases great results have been obtained.

In \cite{Ghinculov} a semi-numerical approach to the calculation of two-loop 
Feynman diagrams was proposed and applied to the two-loop self-energy of
the Higgs boson and to the decay $Z \to b \bar{b}$; 
in \cite{Passarino} a method based on the Bernstein-Tkachov 
theorem was proposed and applied both to multi-leg one-loop and to two-point 
and three-point two-loop Feynman diagrams. In \cite{BinHein}
a numerical method based on the sector decomposition was proposed and
applied to multi-leg one-loop Feynman diagrams calculations as well as to
two-loop and three-loop two-, three- and four-point functions in the 
non-physical region and to the evaluation of phase-space integrals.
Finally, in \cite{Caffo} the numerical evaluation of two-point functions 
was made by means of differential equations solved with the Runge-Kutta
method.

For what concerns the analytical approach, the evaluation of vacuum diagrams,
two-point, three-point and four-point functions (see for example \cite{Twopoint},
\cite{Threepoint,GENSYM,UgoRo1,1dHPLext}, and \cite{Fourpoint,LI,POL2}), used 
also in the case of phase-space integrals (see \cite{Phase}), was made, in the 
last few years, with a variety of different techniques. Since \cite{Chet}, the 
most used one consists in the reduction of the Feynman diagrams to a small set 
of scalar integrals, via integration-by-parts, Lorentz-invariance \cite{LI} and 
general symmetry identities \cite{GENSYM}, followed by the calculation of the 
scalar integrals with different methods such as, for example, the expansion by 
regions \cite{REGIONS}, the Mellin-Barnes transformations \cite{MB}, the 
relations among integrals of different dimension $D$ \cite{Tarasov}, or the 
differential equations method \cite{DiffEq}.

In this paper we will review the algorithm for the reduction of the
Feynman diagrams to the set of independent scalar integrals, called Master
Integrals (MIs) and the differential equations technique for their
evaluation. The problem of the choice of the basis of functions used
for the expression of the analytical results will be also discussed,
giving particular emphasis to the Harmonic Polylogarithms and the
several extensions that took place recently.

\section{Algebraic Reduction to Master Integrals}

The calculation of a physical observable for a certain reaction in 
perturbation theory is connected to the evaluation of the Feynman diagrams
involved in the process. Once the observable is written in terms of Feynman
diagrams and the Dirac algebra is performed (that means usually that the 
traces over the Dirac indices are evaluated), we find an expression
which is a combination of (several) scalar integrals, whose ultraviolet and
infrared divergences are regularized within the dimensional regularization 
scheme. The general structure of such integrals at the 2-loop
level is the following:
\be
I = \int {\mathfrak{D}}^Dk_1 {\mathfrak{D}}^Dk_2 \, \frac{S_{i_{1}}^{s_1}  \cdots   
S_{i_{\overline{t}}}^{s_{\overline{t}}}}{{\mathcal D}_{i_{1}}^{n_1} 
\cdots {\mathcal D}_{i_{t}}^{n_t}} \, , 
\label{eq1}
\ee
where $t$ is the number of different denominators ${\mathcal D}_{i}$ 
(this number is called {\it topology} of the integral), $\bar{t}$ is the 
number of independent scalar products $S_{i}$ on the numerator, and 
${\mathfrak{D}}^Dk$ stands for a suitable integration measurement 
(normalization) for the integrals.
\bfig
\bc
\epsfig{file=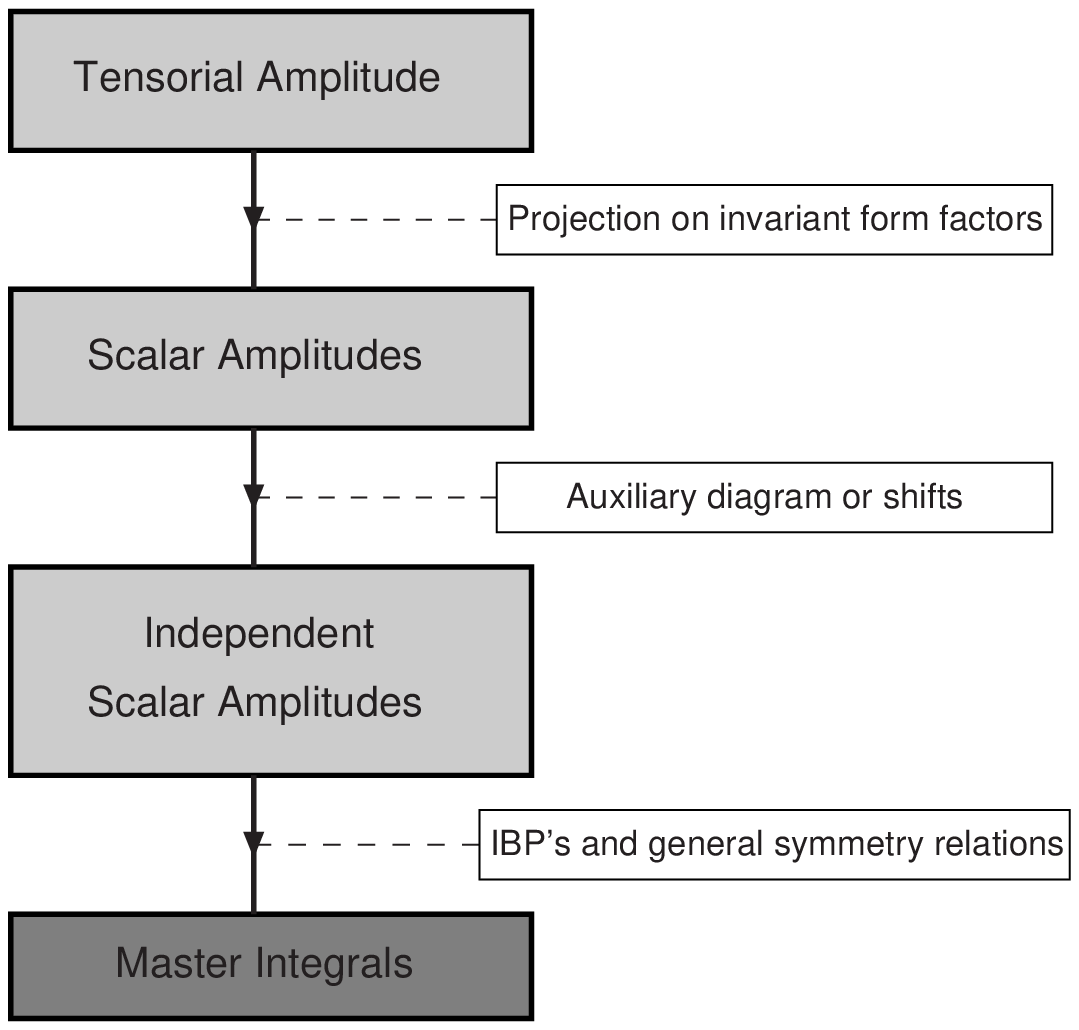,height=7.5cm,width=4.3cm,
        bbllx=230pt,bblly=340pt,bburx=370pt,bbury=660pt}
\ec
\caption{\label{fig1} Flowchart of the method used for the reduction 
to the MIs.}
\efig

The reduction of the integrals of Eq.~(\ref{eq1}) to the MIs is schematically 
represented in the flowchart of Fig.~\ref{fig1} and it is based on the 
following steps:
\begin{enumerate}
\item Once the Feynman diagrams for the evaluation of the observable are
written, we project the corresponding amplitude on a basis of known tensors
in such a way that the amplitude can be written in terms of scalar form
factors. These form factors are expressed in terms of a huge number of
scalar integrals, that can have on the numerator a combination of scalar
products between an external momentum and a loop-momentum or between two
loop-momenta; note that, because we use dimensional regularization, a
simplification between a scalar product and a denominator that contains
this scalar product is possible. Once a denominator disappears, because of
the simplification against a scalar product, the {\it topology} of the 
integral is of course lowered by a unit. In principle, given an integral 
of topology $t$, one must consider all the possible $t!$ subtopologies 
found simplifying repeatedly a denominator against a scalar product on 
the numerator.
\item The scalar integrals are classified with respect to their 
topology. Because we are dealing with integrals that have a suitable mass
distribution (for example we can consider integrals with mass-less
propagators and outgoing legs) it can happen that subtopologies coming from
different simplifications are the same subtopology, but expressed with 
different routings. Therefore, the ``independent'' subtopologies have to 
be chosen and the ``dependent'' subtopologies have to be transformed by 
means of suitable transformations in the independent ones. This analysis
can be done in the framework of the ``Auxiliary diagram scheme'' or the 
``shifts scheme'', as explained extensively in \cite{UgoRo1}.
\item The reduction of the scalar integrals belonging to the independent 
topologies to a ``hopefully'' small set of  master integrals (MIs), is done 
by means of the so-called Laporta algorithm \cite{Lap}, which consists in 
the following. We know that the $D$-regularized scalar integrals coming 
from the projection operation are not all independent, but they satisfy 
certain classes of identity relations. 
\begin{itemize}
\item The most important class is constituted by the integration-by-part 
identities (IBPs), introduced in \cite{Chet}. IBPs link 
scalar integrals of the same topology, but with different power of the
denominator and different scalar products on the numerator, among each other 
and to scalar integrals of subtopologies. IBPs can be written in the following
way:
\be
\int {\mathfrak{D}}^Dk_1 {\mathfrak{D}}^Dk_2 \, \frac{\partial}{
\partial k_{i}^{\mu}} \left\{ v^{\mu} \, \frac{S_{i_1}^{s_{1}} \cdots 
S_{i_{\overline{t}}}^{s_{\overline{t}}}  }  
{{\mathcal D}_{i_1}^{n_{1}} \cdots {\mathcal D}_{i_t}^{n_{t}}}
\right\} = 0  \, ,
\label{eq2}
\ee
where $i=1,2$ and where $v^{\mu} = k_{1}^{\mu},k_{2}^{\mu},p_{i}^{\mu}$ is one 
of the independent vector of the problem. In the case of a 3-point functions,
Eq.~(\ref{eq2}) gives 8 equations for initial scalar amplitude in the brackets. 
For a 4-point function, instead, the IBPs are 10.
\item Another class of identity relations that can be used in the reduction
process is related to the fact that the integrals are Lorentz scalars \cite{LI}.
This property translates into one additional equation for a 3-point function:
\be
\Bigl( p_{1}^{\mu} p_{2}^{\nu} - p_{1}^{\nu} p_{2}^{\mu} \Bigr)
\sum_{n=1}^{2} \left[ p_{n}^{\nu} \frac{\partial }{\partial p_{n}^{
\mu}} - p_{n}^{\mu} \frac{\partial }{\partial 
p_{n}^{\nu}} \right] I(p_i) \, = \, 0 \, ,
\label{LI} 
\ee
where $p_1$ and $p_2$ are the independent vectors of the problem, or 
three additional equations for a 4-point function:
\bea
\Bigl( p_{1}^{\mu} p_{2}^{\nu} - p_{1}^{\nu} p_{2}^{\mu} \Bigr)
\sum_{n=1}^{3} \left[ p_{n}^{\nu} \frac{\partial }{\partial p_{n}^{
\mu}} - p_{n}^{\mu} \frac{\partial }{\partial 
p_{n}^{\nu}} \right] I(p_i) & = & 0 \, , 
\label{LI1} \\
\Bigl( p_{1}^{\mu} p_{3}^{\nu} - p_{1}^{\nu} p_{3}^{\mu} \Bigr)
\sum_{n=1}^{3} \left[ p_{n}^{\nu} \frac{\partial }{\partial p_{n}^{
\mu}} - p_{n}^{\mu} \frac{\partial }{\partial 
p_{n}^{\nu}} \right] I(p_i) & = & 0 \, ,  
\label{LI2} \\
\Bigl( p_{2}^{\mu} p_{3}^{\nu} - p_{2}^{\nu} p_{3}^{\mu} \Bigr)
\sum_{n=1}^{3} \left[ p_{n}^{\nu} \frac{\partial }{\partial p_{n}^{
\mu}} - p_{n}^{\mu} \frac{\partial }{\partial 
p_{n}^{\nu}} \right] I(p_i) & = & 0 \, ,
\label{LI3} 
\eea
where the independent vectors are $p_1$, $p_2$ and $p_3$.
\item In the case in which a suitable mass distribution is considered, we
can find additional equations considering the symmetries of the diagrams
(see \cite{GENSYM}). In general a symmetry of the problem brings to an identity 
that the integrals have to satisfy.
\end{itemize}
Considering all the identity relations mentioned above, a huge system of linear 
equations is constructed. The unknowns are the scalar integrals themselves 
and the point is that the construction of the system can give more equations 
than unknown amplitudes \cite{Lap}, overconstraining formally the system; but 
not all the equations are independent. It can happen that the system is 
effectively overconstrained, and then the solution of the system gives the 
integrals of the topology under consideration as a combination of the MIs of 
the subtopologies, or it is not, and then all the integrals of the topology 
under consideration are expressed as a combination of the MIs of that topology 
(and the MIs of the subtopologies). The solution of the system is performed 
with the Gauss law of substitution. The entire chain is completely algebraic 
and can be implemented in a computer program.
\end{enumerate}

Several authors developed own programs, written in {\tt FORM} \cite{FORM}, 
{\tt C}, or {\tt Mathematica} \cite{MATHE}, for the generation of the linear 
system and for its solution \cite{SOLVE,IdSolver}.
Recently, a computer program written for Maple \cite{MAPLE}, using the Laporta 
algorithm for the reduction of scalar integrals to the MIs was published 
\cite{Anastasiou:2004vj}.

\section{The Differential Equations Method}

The MIs are functions of the external kinematical invariants and, therefore, 
they satisfy a system of first-order linear differential equations. As a simple
example, consider the case in which a topology $t$ has a single MI. Let us 
choose the basic scalar integral of the topology\footnote{The choice of the
set of MIs to which all the scalar integrals belonging to a certain topology 
can be reduced is totally free. One choice is connected to another by the
identity relations constructed for the reduction process. A criterion in the 
preference of one set with respect to the others is, of course, related to the 
solution of the system of differential equations. We choose the set that
satisfy the easiest possible system. That means, usually, the set for which
the system is decoupled exactly in $D$ or at least triangularizes in the limit
$D \to 4$.}:
\be
F(s_{i}) = \int {\mathfrak{D}}^Dk_1 {\mathfrak{D}}^Dk_2 \, 
\frac{1}{{\mathcal D}_{i_1} \cdots {\mathcal D}_{i_t}} \, ,
\ee
where $s_{i}$ stand for the independent invariants that can be constructed
with the external momenta of the problem (for example $s_{1}=p_1^2$,
$s_{2}=p_1 \cdot p_2$, etc.).

The following matrix can be constructed:
\be
O_{jk}(s_{i}) = p_{j}^{\mu} \frac{\partial}{\partial p_{k}^{\mu}} F(s_{i}) \,
.
\label{diffeq1}
\ee

As $F(s_{i})$ depends on the invariants $s_{i}$, we have, on one hand:
\be
  O_{jk}(s_{i}) = p_{j}^{\mu} \sum_{\xi} 
          \frac{\partial s_{\xi}}{\partial p_{k} ^{\mu}} 
          \frac{\partial }{\partial s_{\xi}} F(s_i) 
   =  \sum_{\xi} a_{\xi,jk}(s_{l}) \frac{\partial }{\partial s_{\xi}} 
                                             F(s_i)\, ,
\ee
where the functions $a_{\xi,kl}(s_{i})$ are linear combinations of $s_{i}$.
On the other hand, because $F(s_{i})$ is evaluated in dimensional
regularization, and then it is convergent, we can perform the derivative of
Eq.~(\ref{diffeq1}) directly on the integrand 
$1/{\mathcal D}_{i_1}/../{\mathcal D}_{i_t}$, getting a combination of scalar
integrals with additional power on the denominator and scalar products on the
numerator. Using the identity relations loaded for the reduction to the MIs,
the system of Eq.~(\ref{diffeq1}) becomes a linear system involving the
derivatives of $F(s_{i})$ with respect to the invariants $s_{i}$, $F(s_{i})$
itself and the MIs of the subtopologies. The system can be inverted and the 
differential equations can be written in the following general way:
\be
\frac{\partial }{\partial s_{j}} F(s_i) = A(D,s_i) \, F(s_i) + \Omega(D,s_i) \,
, \label{diffeq2}
\ee
where the non-homogeneous term $\Omega(D,s_i)$ contains the MIs of the
subtopologies, that have to be considered known. The homogeneous term $A(D,s_i)$
gives the analytical structure of the function $F(s_i)$, containing the
thresholds and pseudo-thresholds of the diagram under consideration, that 
appear as singular points for the differential equation\footnote{The 
construction of the system of first-order linear differential equations is 
outlined in the flowchart of Fig.~\ref{fig2}.}. Note that one of the 
differential equations is sufficient for the solution of $F(s_i)$, provided 
that we are able to fix the boundary conditions. Therefore, we can consider 
only the equation with respect to, say, $s$. The solution of Eq.~(\ref{diffeq2})
is done by means of the Euler's method of variation of the arbitrary constants.
\bfig
\bc
\epsfig{file=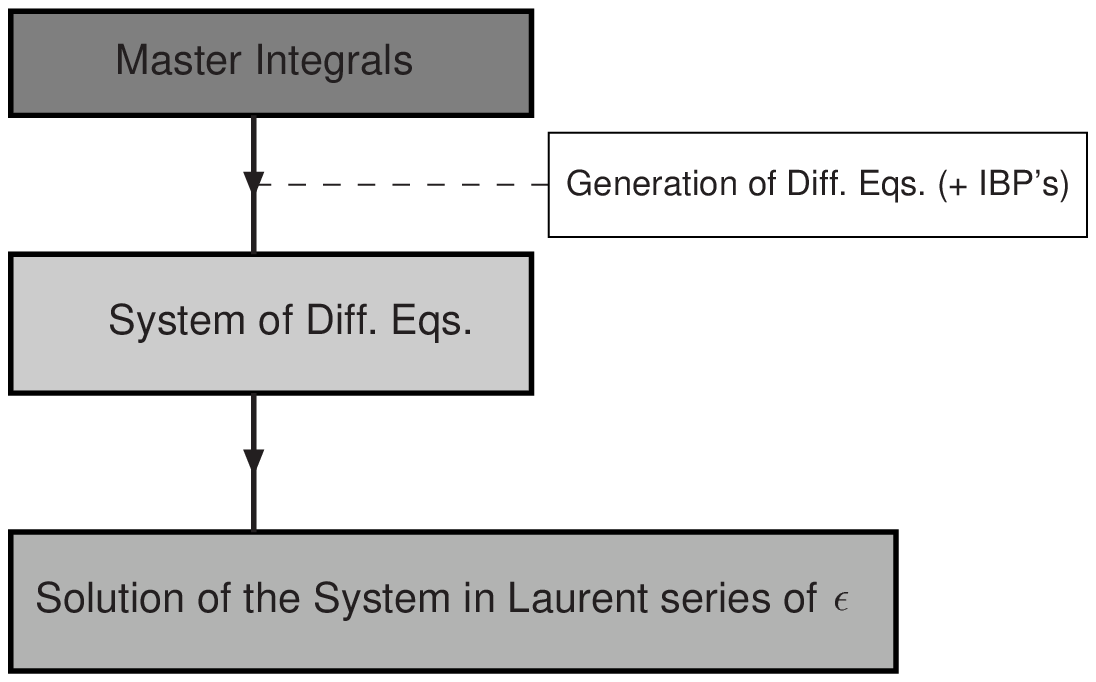,height=5.0cm,width=4.3cm,
        bbllx=230pt,bblly=425pt,bburx=370pt,bbury=630pt}
\ec
\caption{\label{fig2} Flowchart of the Differential Equations method.}
\efig

We can sketch the search for the solution as follows:
\begin{enumerate}
\item We expand $F(s)$ (the dependence on the other invariants is understood) 
in Laurent series of $(D-4)$:
\be
F(s) = \sum_{j=-k}^{n} (D-4)^{j} F_{j}(s) \, ,
\ee
where $k$ is the maximum pole and $n$ is the required order in
$(D-4)$ needed for $F(s)$. Order by order in $(D-4)$ we have, then, to solve 
the equation:
\be
\frac{\partial }{\partial s} F_{j}(s) = A(s) \, F_{j}(s) 
+ \tilde{\Omega}_{j}(s) \, ,
\ee
where for $j=-k$ (the maximum pole) we have 
$\tilde{\Omega}_{-k}(s)=\Omega_{-k}(s)$ ($\Omega(s)$ is the non-homogeneous part
in Eq.~(\ref{diffeq2})) and for $j>-k$, $\tilde{\Omega}_{-k}(s)$
can involve also the previous orders in $(D-4)$ of $F(s)$.
Note that, while the non-homogeneous term $\tilde{\Omega}_{j}(s)$ is different 
at each order in $(D-4)$, the homogeneous equation is the same.
\item We solve the homogeneous equation 
\be
\frac{\partial }{\partial s} f(s) = A(s) \, f(s) \, ,
\ee
finding the solution:
\be
f(s) = \mbox{exp} \left[ \int^{s} A(t) dt \right] \, .
\ee
\item We express the solution of the non-homogeneous equation order by order 
in $(D-4)$ in the integral form:
\be
F_{j}(s) = f(s) \left[ \int^{s} \frac{1}{f(t)} \tilde{\Omega}_{j}(t) dt + k_{j} 
\right] \, , 
\label{repeatint}
\ee
where $k_{j}$ is the arbitrary constant of integration.
\item We fix the constant of integration imposing the initial conditions. In
order to find the initial conditions we have to know additional pieces of 
information about the integral we are calculating. For example it is sufficient 
to know that the integral is regular for some value of $s$.
\end{enumerate}

In the case of $N$ MIs ($N>1$) we have still a system of $N$ first-order 
coupled linear differential equations for every variable $s_i$. As in the
previous case the equations in one of the invariants are sufficient for 
the solution. 

In spite of the elegance of the method, two problems arise.

One is connected with the number of MIs that a topology can have. In fact, 
while the solution of a first-order linear differential equation (case with
one MI) is relatively trivial, a second-order differential equation (case  with
two MIs) can give more problems and starting from the third-order one it can 
be hard to find the solution, except in very particular cases.
We can understand, therefore, the importance of the choice of the set of MIs.
A choice that, even in the case of two or more MIs, could triangularize the
system, at least in the $(D-4)$ expansion, would be of course more suitable than
another; but this choice can not always be done and, in general, there is not 
always a solution for the problem of a topology with many MIs.

Another problem concerns the choice of the basis of functions used for the
expression of the analytical results. Even in this case, there is not a unique
solution, but, nevertheless, one can follow some guidelines. For example
the uniqueness of the representation of the result in terms of these functions; 
the non-redundancy of the representation and the absence of ``hidden zeroes'' 
(that means to avoid representations that can give expressions that are not 
manifestly zero, but, because of the fact that these functions satisfy certain 
relations, they are effectively zero); the total control on the expansions in
all the points of the domain and on the analytical continuation (these two 
``properties'' are needed for a numerical evaluation of the functions and then
of the result).

All these properties are fulfilled, by construction, by the set of functions
called Harmonic Polylogarithms, that we are going to review briefly in the
next paragraph.

\section{Harmonic Polylogarithms and Related Generalized Functions}

In many cases, a suitable integral representation for the Feynman diagrams 
can be found in terms of hypergeometric functions. Nevertheless, in problems
involving many scales, the dependence of the generalized hypergeometric 
functions on these scales is highly non-trivial. Moreover, if 
we regularize the divergent integrals in dimensional regularization, an 
expansion in $(D-4)$ is required, for renormalization purposes. The expansion 
of a hypergeometric function in its parameters can be very complicated 
\cite{Lozano}. This is the reason why one looks for a solution of the 
differential equations directly expanded in $(D-4)$. Order by order in $(D-4)$, 
we can find a representation for the MIs in terms of Nielsen's polylogarithms 
and related functions \cite{Nielsen}. But, firstly, the representation in terms 
of polylogarithms suffers from the problem of ``hidden zeroes'' discussed at 
the end of the previous paragraph; moreover, it can happen that one needs an 
integral representation that does not belong to this class of functions.

One of the most elegant solutions for these two problems is the introduction 
of a basis of transcendental functions defined by repeated integration over 
a set of basic simple functions. Let us consider for example a case in which 
we have at most two different scales in the problem, say $s$ and $a$, in such 
a way that we can construct the dimensionless parameter $x=s/a$; moreover, the 
structure of the thresholds of the Feynman diagrams involved in the calculation 
is such that the possible singular points are $x=0$ and $x=1$ and no squared 
roots are present in the homogeneous equation for the MIs (this is actually 
the case of the majority of the calculations in 
\cite{Twopoint,Threepoint,GENSYM,UgoRo1,1dHPLext}). 
In this situation a suitable basis of functions in
which the results can be expressed are the 1-dimensional harmonic polylogarithms
(HPLs) introduced in \cite{1dHPL}. We consider the following set of three basis 
functions:
\be
g(0;x) = \frac{1}{x} \, , \quad g(1;x) = \frac{1}{1-x} \, , \quad
g(-1;x) = \frac{1}{1+x} \, ,
\label{basis}
\ee
we define the HPLs of weight 1 as:
\bea
H(0,x) & = &  \int_{1}^{x} \frac{dt}{t} = \log{x} \, ,  \\
H(1,x) & = &  \int_{0}^{x} \frac{dt}{1-t} = - \log{(1-x)} \, , \\
H(-1,x) & = & \int_{0}^{x} \frac{dt}{1+t} = \log{(1+x)} \, , 
\eea
and we iterate the previous definition introducing the HPLs of weight $w+1$
\be
H(\vec{0}_{w+1};x) = \frac{1}{(w+1)!}\log^{w+1} x \, , \quad
H(a,\vec{w};x) = \int_0^x f(a;x') H(\vec{w},x'), 
\label{pesi}
\ee
where $a$ can take the values -1, 0 and 1 and $\vec{w}$ is a vector with
$w$ components, consisting of a sequence of -1, 0 and 1 as well.

The set of functions so defined satisfy two important properties: {\it i.}
they form a shuffle algebra, in which the product of two HPLs of weight $w_1$
and $w_2$ is a HPL of weight $w_1+w_2$; {\it ii.} they form a closed set 
under certain transformations of the argument. In particular, this
property is needed for the asymptotic evaluation of such a functions,
connecting the expansion in $x \to 0$ to the one in $x \to \infty$.
Another important property of HPLs is that their analytical structure is 
manifestly shown. The integral representations of Eq.~(\ref{repeatint}) can 
be expressed in terms of HPLs directly or re-conducted to HPLs via simple 
integration by parts.

The choice of the basis (\ref{basis}) for the expression of the HPLs is 
connected with the
structure of the thresholds of the diagrams involved in the calculation. It 
can happen that the solution of the homogeneous differential equation contains a
square root, that can not be avoided by a suitable change of variables. 
Moreover, we can deal with a calculation in which, for example, the diagrams 
can have three different kind of thresholds: $x=0$, $x=1$ and $x=4$. In this 
case an extension of the basis of HPLs is needed \cite{1dHPLext}. 
We introduce the following additional basis functions:
\bea
& & g(\mp 4;x) = \frac{1}{4 \pm x} \, , \quad
g(c,x) = \frac{1}{ x-\frac{1}{2}-i \frac{\sqrt{3}}{2} } \, ,  \quad
g(\overline{c},x) = \frac{1}{ x-\frac{1}{2}+i \frac{\sqrt{3}}{2} }  \nn\\
& & g(\mp r,x) = \frac{1}{\sqrt{x(4 \pm x)}}  \, , \quad
g(\mp 1 \mp r,x) = \frac{1}{\sqrt{x(4 \pm x)}~(1 \pm x)} \, ,\nn\\
& & g(\pm 1/4;x) = \frac{1}{\frac{1}{4} \mp x}  \, , \quad
g(\pm 1 \pm r/4;x) = \frac{1}{\sqrt{x \mp \frac{1}{4}} (1 \mp x)}  \, ,\nn\\
& & g(r_0/4;x) = \frac{1-2i \sqrt{x-\frac{1}{4}}}{x \sqrt{x-\frac{1}{4}}}  
\, , \quad
g(-r_0/4;x) = \frac{1-2 \sqrt{x+\frac{1}{4}}}{x \sqrt{x+\frac{1}{4}}} \, .
\label{basisext}
\eea

The relative weight-1 HPLs are defined as the integration between 0 and $x$ of
the previous basis functions (all the functions in Eq.~(\ref{basisext}) are
integrable in $x=0$) and the generalization to higher weights is done by
Eq.~(\ref{pesi}) in which now the single weight can take the values
0, $\pm 1$, $\pm 4$, $\pm r$, $c$, $\bar{c}$, $\mp 1 \mp r$, $\pm 1/4$,
$\pm 1 \pm r/4$, $\pm r_0/4$.

The set of generalized HPLs so introduced satisfy all the properties of the
former set of HPLs. Similar generalizations can be carried out in the case 
of different thresholds.

Another generalization can be done in the case in which we have three scales
in the game and, therefore, two dimensionless parameters can be constructed.
This gives the so-called 2-dimensional HPLs introduced in \cite{2dHPL,POL2} 
and their further extension introduced in \cite{2dHPLext}.

As a remark, note that the representation of the analytical results in terms 
of HPLs allows a perfect control on the numerical evaluations.

\section{Summary}

In this paper the Laporta algorithm for the reduction of the Feynman diagrams 
to the master integrals and the differential equations technique for their
analytical evaluation are reviewed. A particular attention is paid to the
problem of the choice of the basis of functions in which the result can be
expressed. Some argument for the use of HPLs is done. In particular, the
possibility of an extension of the set of functions to the case of problems
involving different thresholds and different scales (more than two) is 
briefly outlined.


\end{document}